# Distributed analysis of nonlinear wave mixing in fiber due to forward Brillouin scattering and Kerr effects

YOSEF LONDON, HILEL HAGAI DIAMANDI[a], GIL BASHAN, AND AVI ZADOK[*]

Faculty of Engineering and Institute of Nano-Technology and Advanced Materials, Bar-Ilan University, Ramat-Gan 5290002, Israel

**Abstract**

Forward stimulated Brillouin scattering (F-SBS) is a third-order nonlinear-optical mechanism that couples between two co-propagating optical fields and a guided acoustic mode in a common medium. F-SBS gives rise to nonlinear wave mixing along optical fibers, which adds up with four-wave mixing induced by the Kerr effect. In this work, we report the distributed mapping of nonlinear wave mixing processes involving both mechanisms along standard single-mode fiber, in analysis, simulation and experiment. Measurements are based on a multi-tone, optical time-domain reflectometry setup, which is highly frequency-selective. The results show that F-SBS leads to nonlinear wave mixing processes that are more complex than those that are driven by the Kerr effect alone. The dynamics are strongly dependent on the exact frequency detuning between optical field components. When the detuning is chosen near an F-SBS resonance, the process becomes asymmetric. Power is coupled from an upper-frequency input pump wave to a lower-frequency one, and the amplification of Stokes-wave sidebands is more pronounced than that of anti-Stokes-wave sidebands. The results are applicable to a new class of distributed fiber-optic sensors, based on F-SBS.

[a] Y. London and H. H. Diamandi contributed equally to this work
[*] Corresponding author: Avinoam.Zadok@biu.ac.il

**Main Text**

**1. Introduction**

Forward stimulated Brillouin scattering (F-SBS) is a nonlinear optical interaction between two co-propagating optical field components and a guided acoustic wave within an optical medium [1-3]. The phenomenon has been observed in standard optical fibers since the 1980's [1-6], and has been studied extensively in photonic-crystal [7-12], micro-structured [13-16] and tapered fibers [17-18]. Striking demonstrations of F-SBS in silicon-photonic integrated devices have been reported as well [19-21]. Interest in F-SBS in standard fiber has been reawakened recently: the mechanism was shown to support the fiber-optic sensing of liquid media outside the cladding of standard, unmodified fiber, where light cannot reach [22]. Several additional reports successfully followed the initial demonstration [23-25]. F-SBS can also introduce inter-core, opto-mechanical cross-phase modulation among the constituent cores of a multi-core fiber [26], and may lead to electro-opto-mechanical oscillations in standard and multi-core fibers [27-28].

F-SBS is a third-order nonlinear-optical mechanism. As such, it may give rise to nonlinear wave mixing processes, through an acoustic idler wave. Nonlinear propagation induced by F-SBS in optical fibers adds on top of four-wave mixing (FWM) due to the Kerr effect, which is widely considered [29]. Wang and coworkers [6], and later Chow *et al.* [24], observed the spectra of nonlinear wave mixing due to combined F-SBS and Kerr effects at the output of optical fibers. Both works observed spectra that are more complex than those obtained through Kerr FWM alone. However, these works did not include distributed mapping. Spatially-distributed analysis of FWM in fiber has been carried out, based on the measurements of Rayleigh scattering [30] and backwards stimulated Brillouin scattering (B-SBS) [31]. Both studies addressed FWM processes due to the Kerr effect only. Distributed analysis of F-SBS has also been reported over a short segment of a nano-structured fiber, using vibrometric measurements from outside the fiber [15]. Spatially-continuous, nonlinear propagation involving F-SBS is addressed in a recent theoretical study by Wolff and coworkers [32]. However, a corresponding experimental characterization has not yet been reported.

In this work we report an experimental distributed analysis of nonlinear wave mixing due to both F-SBS and the Kerr effect along 8 km of standard single-mode fiber. The measurements are based on a multi-tone, optical time-domain reflectometry (multi-tone OTDR) setup: the monitoring of Rayleigh back-scatter of multiple spectral field components as a function of time [30]. The extension of this measurement principle to address F-SBS presents two main challenges. First, the separation of back-scattered components that are detuned in optical frequency by only several hundreds of MHz is necessary. Such separation is obtained here using tunable, narrowband B-SBS amplification [33]. Second, the efficient stimulation of F-SBS requires input optical fields that are highly coherent, whereas Rayleigh back-scatter of coherent light is extremely noisy [34-35]. This difficulty is resolved in the experimental setup as well.

The results show that wave mixing involving F-SBS is highly sensitive to the exact frequency offset between incident optical fields. When the offset matches a resonance frequency of F-SBS, the process becomes asymmetrical: Power is coupled from a higher-frequency input pump wave to a lower-frequency one, and the Stokes-wave sidebands are amplified more efficiently than their anti-Stokes wave counterparts. This asymmetry is in marked contrast to the symmetric characteristics of FWM due to Kerr nonlinearity. In addition, detuning of the frequency offset between the incident waves above or below the F-SBS resonance gives rise to nonlinear mixing processes that are qualitatively different. Measurements are in agreement with analysis and simulations.

The observations contribute a more complete description of nonlinear wave mixing in fiber, for a more general case where the Kerr effect is not the exclusive underlying mechanism. On top of the interest to basic research, the results are also relevant to emerging new concepts of distributed fiber-optic sensing that are based on F-SBS [36-37]. These measurement schemes allow for the analysis of media outside the fiber, where light cannot reach [36-37]. However, the signal-to-noise ratio, range and resolution of at least one such protocol are currently limited by the onset of nonlinear amplification of modulation sidebands [36]. A more complete description of wave mixing involving F-SBS may be incorporated in the sensor arrangement, and improve its performance. The study is therefore of timely and practical interest.

A model for nonlinear wave mixing through the combination of F-SBS and the Kerr effect is presented in Section 2. Approximate analytic solutions are proposed, and a numerical analysis is reported. Experimental results are provided in Section 3, and a summary is given in Section 4. Preliminary results were briefly reported in conference proceedings [38,39].

## 2. Analysis and simulation of nonlinear wave mixing

### *2.1 Forward stimulated Brillouin scattering*

We begin with a brief introduction of F-SBS. Detailed analysis and discussion of the effect may be found in numerous earlier works [1-6,26]. In addition to the single optical mode, standard fibers also support a variety of guided mechanical (acoustic) modes that propagate along the fiber axis. This study is restricted to radial modes, denoted by $R_{0,m}$ where $m$ is an integer, in which the transverse displacement of the fiber material is purely radial [1-3]. The propagation of each mode is characterized by a cut-off frequency $\Omega_m$ [1-3]. Close to cut-off, the axial phase velocities of the guided acoustic modes approach infinity. Hence for each mode $R_{0,m}$ there exists a frequency, very near cut-off, for which the axial phase velocity matches that of the optical mode [1-3]. Two co-propagating optical field components that are offset in frequency by $\Omega \approx \Omega_m$ may be coupled with the guided acoustic mode [1-3]. The optical fields may stimulate the acoustic wave via electrostriction. The acoustic wave, in turn, can scatter and modulate light through photo-elasticity.

Spontaneous scattering by acoustic modes is often denoted as guided acoustic waves Brillouin scattering [1]. The stimulated effect, in which the acoustic wave oscillations are driven by light, is referred to as F-SBS. The efficiency of stimulated scattering through a given acoustic mode is determined by spatial overlap considerations (see also below, [1-3,26]). The magnitude of F-SBS may be quantified in terms of an equivalent opto-mechanical nonlinear coefficient $\gamma_{\text{OM}}^{(m)}(\Omega)$, in units of [W×m]$^{-1}$ [16,26]. The coefficient is analogous to that associated with Kerr nonlinearity, $\gamma_{\text{Kerr}}$.

The opto-mechanical nonlinear coefficient is strongly dependent on the exact frequency detuning $\Omega$ between the pair of stimulating optical fields. The F-SBS spectra are of Lorentzian line-shapes, with local maximum values $\gamma_{\text{OM}}^{(0,m)}$ at resonance frequencies $\Omega = \Omega_m$, and linewidths $\Gamma_m$ [16,26]. F-SBS in standard single-mode fibers with 125 µm cladding diameter and standard, dual-layer acrylate coating is the most efficient through radial mode $R_{0,7}$, with a resonance frequency near 320 MHz, a linewidth of about 4-5 MHz, and a maximal opto-mechanical nonlinear coefficient $\gamma_{\text{OM}}^{(0,7)}$ of about 2 [W×km]$^{-1}$ [27]. Note that the opto-mechanical nonlinear coefficient on resonance is comparable with typical values of $\gamma_{\text{Kerr}}$.

*2.2 Nonlinear polarization terms, nonlinear wave equations and approximate solutions*

Consider two continuous optical field components of frequencies $\omega_l = \omega + \left(l - \tfrac{1}{2}\right)\Omega$, where $\omega$ denotes a central optical frequency, $\Omega$ represents a radio-frequency offset, and $l = 0,1$. The two fields co-propagate in a standard optical fiber, in the positive $\hat{\mathbf{z}}$ direction. Let us denote the complex Jones vectors of the two wave components as $\vec{E}_l(z)$, where $z$ represents axial position. The optical waves may stimulate the oscillations of a guided acoustic mode $R_{0,m}$, at frequency $\Omega$. One may show that the complex magnitude of the material displacement of the stimulated acoustic wave is given by [36]:

$$B^{(m)}(z) = -\frac{\varepsilon_0}{4\rho_0\Omega_m\Gamma_m} Q_{ES}^{(m)} \frac{1}{j+\Delta} \vec{E}_0^\dagger(z)\vec{E}_1(z). \tag{1}$$

Here $\rho_0$ is the density of silica, $\varepsilon_0$ is the vacuum permittivity, $\Delta \equiv 2(\Omega_m - \Omega)/\Gamma_m$, and $Q_{ES}^{(m)}$ denotes the spatial overlap integral between the transverse profile of the electrostrictive driving force and that of the modal acoustic displacement [26]. The strain associated with the acoustic wave induces nonlinear polarization, through photo-elasticity. The vector magnitudes of the nonlinear polarization components at frequencies $\omega_{1,0}$ are given by [36]:

$$\vec{P}_{NL,1}^{(m)}(z) = \varepsilon_0 Q_{PE}^{(m)} B^{(m)}(z)\vec{E}_0(z) = -\frac{\varepsilon_0^2}{4\rho_0\Omega_m\Gamma_m} Q_{ES}^{(m)} Q_{PE}^{(m)} \frac{1}{j+\Delta}\left[\vec{E}_0^\dagger(z)\vec{E}_1(z)\right]\vec{E}_0(z), \tag{2}$$

$$\vec{P}_{NL,0}^{(m)}(z) = \varepsilon_0 Q_{PE}^{(m)} \left[B^{(m)}(z)\right]^* \vec{E}_1(z) = \frac{\varepsilon_0^2}{4\rho_0\Omega_m\Gamma_m} Q_{ES}^{(m)} Q_{PE}^{(m)} \frac{1}{j-\Delta}\left[\vec{E}_1^\dagger(z)\vec{E}_0(z)\right]\vec{E}_1(z). \tag{3}$$

In Eq. (2) and Eq. (3), $Q_{PE}^{(m)}$ is the overlap integral between the transverse profile of photo-elastic perturbation of the dielectric constant due to $R_{0,m}$, and that of the optical mode. Using the slowly varying envelope approximation, the coupled nonlinear wave equations for the complex envelopes of the two fields may be brought to the following form [36]:

$$\frac{d\vec{E}_1(z)}{dz} = -\frac{k_0\varepsilon_0 Q_{ES}^{(m)} Q_{PE}^{(m)}}{8n_0\rho_0\Omega_m\Gamma_m} \times \frac{1-j\Delta}{1+\Delta^2}\left[\vec{E}_0^\dagger(z)\vec{E}_1(z)\right]\vec{E}_0(z), \tag{4}$$

$$\frac{d\vec{E}_0(z)}{dz} = \frac{k_0\varepsilon_0 Q_{ES}^{(m)} Q_{PE}^{(m)}}{8n_0\rho_0\Omega_m\Gamma_m} \times \frac{1+j\Delta}{1+\Delta^2}\left[\vec{E}_1^\dagger(z)\vec{E}_0(z)\right]\vec{E}_1(z). \tag{5}$$

Here $n_0$ is the refractive index of silica and $k_0$ denotes the vacuum wavenumber at the central optical frequency $\omega$. Let us scale the field magnitudes: $\vec{A}_l \equiv \sqrt{2n_0 c\varepsilon_0}\vec{E}_l$ where $c$ is the speed of light in vacuum, so that the optical power levels (in [W]) of the two field components may be written as $P_l = \vec{A}_l^\dagger \vec{A}_l$. With this change of variables, the nonlinear wave equations become:

$$\frac{d\vec{A}_1(z)}{dz} = -\frac{k_0 Q_{ES}^{(m)} Q_{PE}^{(m)}}{16 n_0^2 c \rho_0 \Omega_m \Gamma_m} \times \frac{1-j\Delta}{1+\Delta^2}\left[\vec{A}_0^\dagger(z)\vec{A}_1(z)\right]\vec{A}_0(z) = -\frac{1}{2}\gamma_{OM}^{(0,m)}\frac{1-j\Delta}{1+\Delta^2}\left[\vec{A}_0^\dagger(z)\vec{A}_1(z)\right]\vec{A}_0(z), \tag{6}$$

$$\frac{d\vec{A}_0(z)}{dz} = \frac{k_0 Q_{ES}^{(m)} Q_{PE}^{(m)}}{16 n_0^2 c \rho_0 \Omega_m \Gamma_m} \times \frac{1-j\Delta}{1+\Delta^2}\left[\vec{A}_1^\dagger(z)\vec{A}_0(z)\right]\vec{A}_1(z) = \frac{1}{2}\gamma_{OM}^{(0,m)}\frac{1+j\Delta}{1+\Delta^2}\left[\vec{A}_1^\dagger(z)\vec{A}_0(z)\right]\vec{A}_1(z). \tag{7}$$

In Eq. (6) and Eq. (7), the opto-mechanical nonlinear coefficient $\gamma_{OM}^{(0,m)}$ has been defined [16,26]:

$$\gamma_{OM}^{(0,m)} \equiv \frac{k_0 Q_{ES}^{(m)} Q_{PE}^{(m)}}{8 n_0^2 c \rho_0 \Omega_m \Gamma_m}. \tag{8}$$

The units of the nonlinear coefficient are [W×m]$^{-1}$. The pair of equations may be readily extended to include the Kerr effect, and linear losses with a coefficient $\alpha$ [40-42]:

$$\frac{d\vec{A}_1}{dz} = -\frac{\alpha}{2}\vec{A}_1 + j\frac{8}{9}\gamma_{Kerr}\left[\left(\vec{A}_0^\dagger \vec{A}_0\right)\vec{A}_1 + \left(\vec{A}_1^\dagger \vec{A}_1\right)\vec{A}_1 + \left(\vec{A}_0^\dagger \vec{A}_1\right)\vec{A}_0\right] - \frac{1}{2}\gamma_{OM}^{(0,m)}\frac{1-j\Delta}{1+\Delta^2}\left(\vec{A}_0^\dagger \vec{A}_1\right)\vec{A}_0, \tag{9}$$

$$\frac{d\vec{A}_0(z)}{dz} = -\frac{\alpha}{2}\vec{A}_0 + j\frac{8}{9}\gamma_{Kerr}\left[\left(\vec{A}_0^\dagger \vec{A}_0\right)\vec{A}_0 + \left(\vec{A}_1^\dagger \vec{A}_1\right)\vec{A}_0 + \left(\vec{A}_1^\dagger \vec{A}_0\right)\vec{A}_1\right] + \frac{1}{2}\gamma_{OM}^{(0,m)}\frac{1+j\Delta}{1+\Delta^2}\left(\vec{A}_1^\dagger \vec{A}_0\right)\vec{A}_1. \tag{10}$$

It is assumed herein that the length scale of nonlinear wave mixing processes is much longer than the beat length of residual linear birefringence in standard fiber [40-42]. The chromatic dispersion of standard fibers is not considered. It has negligible effect for the values of $\Omega$ and lengths of fiber used in this work.

Both F-SBS and the Kerr effect are third-order nonlinear-optical phenomena. As such, they both bring about nonlinear mixing between the two optical waves. However, there are several differences between the two contributions. The Kerr nonlinearity is parametric: it preserves the overall energy of the optical field terms combined. F-SBS, on the other hand, is a dissipative effect, in which part of the optical energy is lost to the acoustic wave stimulation. The Kerr terms describe FWM behavior (which is degenerate when only two optical components are considered), with no additional waves involved in the medium. In contrast, mixing between optical waves through F-SBS is mediated by an additional idler, in the form of the acoustic wave. Hence we do not refer to the F-SBS terms strictly as those of a FWM process.

Further, the Kerr effect is independent of frequency changes on the scale of $\Omega$, whereas F-SBS is extremely frequency-selective. As seen in Eq. (2) and Eq. (3), the nonlinear polarization terms due to F-SBS exhibit the resonance response of a second-order system. This response adds a $\pi/2$ phase shift on resonance. The phase shift, in turn, manifests in Eq. (9) and Eq. (10) in nonlinear terms with non-zero real parts, as opposed to those of the Kerr effect that are purely imaginary. Resonances associated with the Kerr effect occur at frequencies outside the optical range. The primary objective of this work is to study the interplay between Kerr and F-SBS terms in nonlinear wave mixing, through analysis, numerical calculations and experiments.

For two input fields that are co-polarized, analytic solutions for the power levels $P_{1,0}(z)$ exist in the following form [29]:

$$P_1(z) = P_1(0) \frac{1+M^{-1}}{1+M^{-1} \exp\left[g^{(m)}(\Delta) L_{\text{eff}}\right]} \exp(-\alpha z), \tag{11}$$

$$P_0(z) = P_0(0) \frac{1+M}{1+M \exp\left[-g^{(m)}(\Delta) L_{\text{eff}}\right]} \exp(-\alpha z). \tag{12}$$

Here $L_{\text{eff}} \equiv \left[1-\exp(-\alpha z)\right]/\alpha$ is the effective length up to point $z$, $M \equiv P_1(0)/P_0(0)$ denotes the ratio between input power levels, and the gain coefficient in units of m$^{-1}$ is defined by:

$$g^{(m)}(\Delta) \equiv P_1(0)(1+M^{-1})\frac{\gamma_{\text{OM}}^{(0,m)}}{1+\Delta^2} \tag{13}$$

The results show that the stimulation of the acoustic wave is associated with the coupling of power from the higher-frequency optical component to the lower frequency one, as may be expected. In addition to $\omega_{0,1}$, however, the stimulated acoustic wave also induces nonlinear polarization components at additional frequencies, $\omega_{2,-1} = \omega \pm \tfrac{3}{2}\Omega$, with respective vector magnitudes:

$$\vec{P}_{\text{NL},2}^{(m)}(z) = \varepsilon_0 Q_{\text{PE}}^{(m)} B^{(m)}(z)\vec{E}_1(z) = -\frac{\varepsilon_0^2}{4\rho_0 \Omega_m \Gamma_m} Q_{\text{ES}}^{(m)} Q_{\text{PE}}^{(m)} \frac{1}{j+\Delta}\left[\vec{E}_0^\dagger(z)\vec{E}_1(z)\right]\vec{E}_1(z), \tag{14}$$

$$\vec{P}_{\text{NL},-1}^{(m)}(z) = \varepsilon_0 Q_{\text{PE}}^{(m)} \left[B^{(m)}(z)\right]^* \vec{E}_0(z) = \frac{\varepsilon_0^2}{4\rho_0 \Omega_m \Gamma_m} Q_{\text{ES}}^{(m)} Q_{\text{PE}}^{(m)} \frac{1}{j-\Delta}\left[\vec{E}_1^\dagger(z)\vec{E}_0(z)\right]\vec{E}_0(z). \tag{15}$$

These nonlinear polarization terms add up with those due to the Kerr effect, and may contribute to the generation of anti-Stokes-wave and Stokes-wave sidebands at optical frequencies $\omega_{2,-1}$. The processes may be described in terms of the following nonlinear wave equations:

$$\frac{d\vec{A}_2}{dz} = -\frac{\alpha}{2}\vec{A}_2 + j\frac{8}{9}\gamma_{\text{Kerr}}\left(\vec{A}_0^\dagger \vec{A}_1\right)\vec{A}_1 - \frac{1}{2}\gamma_{\text{OM}}^{(0,m)}\frac{1-j\Delta}{1+\Delta^2}\left(\vec{A}_0^\dagger \vec{A}_1\right)\vec{A}_1, \tag{16}$$

$$\frac{d\vec{A}_{-1}(z)}{dz} = -\frac{\alpha}{2}\vec{A}_{-1} + j\frac{8}{9}\gamma_{\text{Kerr}}\left(\vec{A}_1^\dagger \vec{A}_0\right)\vec{A}_0 + \frac{1}{2}\gamma_{\text{OM}}^{(0,m)}\frac{1+j\Delta}{1+\Delta^2}\left(\vec{A}_1^\dagger \vec{A}_0\right)\vec{A}_0. \tag{17}$$

Here $\vec{A}_{2,-1}(z)$ denote the complex Jones vector envelopes of the field components at frequencies $\omega_{2,-1}$, scaled as above. First-order approximations for the power levels of the two sidebands may be obtained in terms of equivalent overall nonlinear coefficients:

$$\gamma_{\text{Eq},2}(\Delta) \equiv j\frac{8}{9}\gamma_{\text{Kerr}} - \frac{1}{2}\gamma_{\text{OM}}^{(0,m)}\frac{1-j\Delta}{1+\Delta^2}, \tag{18}$$

$$\gamma_{\text{Eq.-1}}(\Delta) \equiv j\frac{8}{9}\gamma_{\text{Kerr}} + \frac{1}{2}\gamma_{\text{OM}}^{(0,m)}\frac{1+j\Delta}{1+\Delta^2}, \tag{19}$$

$$P_2(z) \approx |\gamma_{\text{Eq.2}}(\Delta)|^2 L_{\text{eff}}^2 P_1^2(0) P_0(0) \exp(-\alpha z), \tag{20}$$

$$P_{-1}(z) \approx |\gamma_{\text{Eq.-1}}(\Delta)|^2 L_{\text{eff}}^2 P_0^2(0) P_1(0) \exp(-\alpha z). \tag{21}$$

Here $P_{2,-1}(z)$ represent the local power levels of the anti-Stokes wave and Stoke-wave sidebands, respectively. Due to the coupling of optical power from $P_1(z)$ to $P_0(z)$ (Eq. (11) and Eq. (12)), we may qualitatively predict that the nonlinear amplification of $P_{-1}(z)$ would be more efficient than that of $P_2(z)$. However this trend is not accounted for by the simplified solutions of Eq. (20) and Eq. (21). These first-order expressions are quantitatively relevant, therefore, only as long as coupling between the two input tones remains comparatively modest. In this regime, the sidebands power levels $P_{2,-1}(z)$ are also fairly weak. Note also that Eq. (11) and Eq. (12) themselves are only valid when no other field components exist. When the sidebands power levels $P_{2,-1}(z)$ become sufficiently high, the approximations for $P_{1,0}(z)$ no longer hold either.

Despite the above limitations, the first-order expressions for $P_{2,-1}(z)$ forecast a dependence of the nonlinear wave mixing process on the sign of $\Delta$. Let us assume that the two input fields $\vec{A}_{0,1}(0)$ are co-polarized and of equal phases. As illustrated in Fig. 1, the phasor addition of nonlinear coefficients due to F-SBS and the Kerr effect would be partially constructive for $\Delta > 0$, and partially destructive for $\Delta < 0$. The equivalent nonlinear coefficients for both sidebands are therefore larger for $\Delta > 0$ (see Eq. (18) and Eq. (19)), suggesting a more efficient amplification of sidebands with $\Omega$ below the F-SBS resonance $\Omega_m$. This prediction is tested against numerical simulations and experiment in subsequent sections.

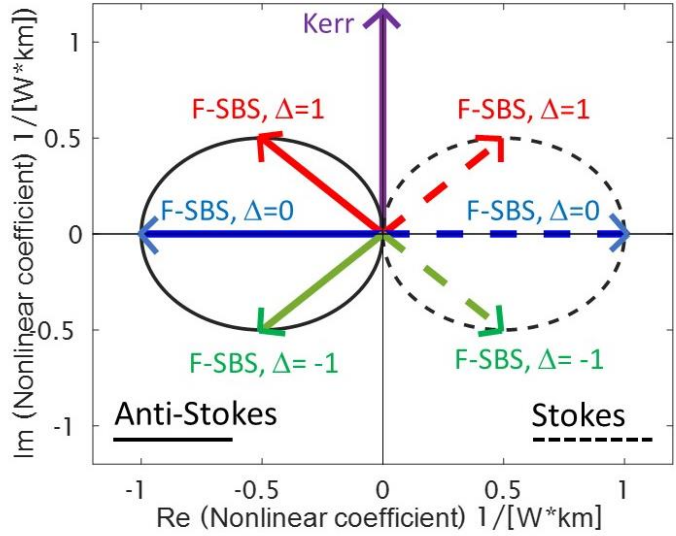

Fig. 1. Phasor diagram of contributions to the equivalent nonlinear coefficients in wave-mixing generation of sidebands (see Eq. (18) and Eq. (19)). Values of $\gamma_{\text{Kerr}} = 1.3$ [W×km]$^{-1}$ and $\gamma_{\text{OM}}^{(0,7)} = 2.0$ [W×km]$^{-1}$ were used. The contribution of the Kerr effect (purple) is purely imaginary and positive. The coefficients due to F-SBS for the Stokes sideband (frequency $\omega_0 - \frac{3}{2}\Omega$, dashed lines) are in the right half-plane. The corresponding coefficients for the anti-Stokes sideband (frequency $\omega_0 + \frac{3}{2}\Omega$, solid lines) are in the left half-plane. On the F-SBS resonance ($\Delta = 0$, blue), both F-SBS coefficients are real-valued. For acoustic frequencies below resonance ($\Delta > 0$), the F-SBS coefficients for both sidebands are in the upper half of the complex plane, representing partially constructive addition of the Kerr and F-SBS terms. This is illustrated for the case of $\Delta = 1$ (red). In contrast, the F-SBS coefficients for $\Delta < 0$ are in the lower half plane, leading to partially destructive phasor summation (illustrated for $\Delta = -1$, green). Black contours mark the range of possible values of the F-SBS contributions, scanning over $\Delta$. The solid (dashed) trace corresponds to the anti-Stokes (Stokes) sideband.

In the general case, nonlinear coupling among all four waves, involving both F-SBS and the Kerr effect, must be considered. Cascaded generation of higher-order sidebands may take place as well. Numerical simulations of the more general problem are described in the next sub-section.

*2.3 Numerical simulations of nonlinear wave mixing*

In this sub-section, the above model is extended to describe nonlinear wave mixing among a larger set of optical field components. Let $\{\vec{A}_n(z)\}$, $n = (-N+1),...,N$ denote the complex Jones vector envelopes of $2N$ continuous optical waves that co-propagate in the positive $\hat{z}$ direction along a standard single-mode fiber. The optical frequencies of the wave components are $\omega_n = \omega + \left(n - \tfrac{1}{2}\right)\Omega$. The frequency offset is chosen near the strongest resonance of F-SBS: $\Omega \approx \Omega_7$. The nonlinear, coupled wave equations presented in the previous sub-section may be generalized to the following set:

$$\frac{d\vec{A}_n}{dz} = -\frac{1}{2}\alpha \vec{A}_n + j\frac{8}{9}\gamma_{\text{Kerr}} \sum_{i,j=(-N+1)}^{N}\left(\vec{A}_j^\dagger \vec{A}_i\right)\vec{A}_{n+j-i} + \frac{1}{2}\gamma_{\text{OM}}^{(0,7)} \sum_{i=(-N+1)}^{N}\left[\frac{1+j\Delta}{1+\Delta^2}\left(\vec{A}_{i+1}^\dagger \vec{A}_i\right)\vec{A}_{n+1} - \frac{1-j\Delta}{1+\Delta^2}\left(\vec{A}_{i-1}^\dagger \vec{A}_i\right)\vec{A}_{n-1}\right]. \tag{22}$$

In principle, optical field components that are detuned by $2\Omega$ might also be coupled through the radial guided mode $R_{0,14}$, which has a cut-off frequency $\Omega_{14} = 2.035\Omega_7$ [1]. However, $\gamma_{\text{OM}}^{(0,14)}$ in standard single-mode fiber is considerably smaller than $\gamma_{\text{OM}}^{(0,7)}$ due to spatial overlap considerations [27]. We verified that the addition of F-SBS through mode $R_{0,14}$ to the above model has marginal effect on the calculated outcome.

| Parameter | Value | Source |
|---|---|---|
| Linear losses coefficient | $\alpha = 0.046$ km$^{-1}$ | OTDR calibration |
| Kerr effect nonlinear coefficient | $\gamma_{\text{Kerr}} = 1.3$ [W×km]$^{-1}$ | Specifications |
| F-SBS nonlinear coefficient | $\gamma_{\text{OM}}^{(0,7)} = 2.0$ [W×km]$^{-1}$ | Calculations [27] |
| Acoustic resonance frequency | $\Omega_7 = 2\pi \times 321.75$ MHz | Measurements, following [22,26] |
| Acoustic modal linewidth | $\Gamma_7 = 2\pi \times 4.3$ MHz | Measurements, following [22,26] |
| Input power, upper-frequency tone | $P_1(0) = 80$ mW | Experimental conditions |
| Input power, lower-frequency tone | $P_0(0) = 80$ mW | Experimental conditions |

Table 1. List of parameters used in numerical simulations of nonlinear wave-mixing.

The local power levels of the optical tones $P_n(z) = \left|\vec{A}_n(z)\right|^2$ were calculated using numerical integration of the above set of equations. The parameters used in the simulations are listed in Table 1. The input phases of the two optical fields were assumed to be equal, and their input states of polarization were aligned. Results are presented for the local power levels of six optical field components as a function of position: $P_k(z)$, $k = -2, -1, 0, 1, 2, 3$. The notation of frequency components used throughout this work is illustrated in Fig. 2, for better clarity. Due to nonlinear coupling, the power levels $P_k(z)$ are also affected by higher-order sidebands. Therefore, an overall number of $2N = 10$ spectral components were used in the numerical calculations. We verified that the addition of further terms beyond 10 has negligible effect on the six $P_k(z)$ traces subject to the specific boundary conditions.

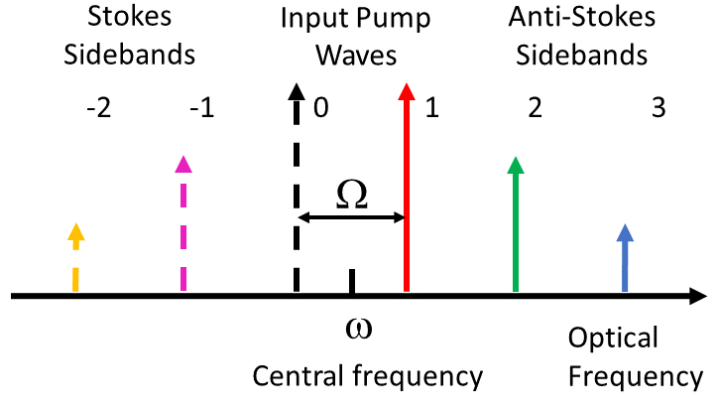

Fig. 2. Illustration and notation of six optical field components at frequencies $\omega + \left(k - \tfrac{1}{2}\right)\Omega$, $k = -2, -1, 0, 1, 2, 3$. Nonlinear wave mixing effects among the field components are simulated and presented in subsequent figures.

Figure 3(a) shows the calculated $P_k(z)$ with a frequency offset $\Omega$ off the F-SBS resonance ($\Delta = 8$). The power levels of the two input wave components are indistinguishable along the entire fiber, as expected. Generation of Stokes-wave and anti-Stokes-wave sidebands is observed. The amplification of the two sidebands along the fiber is symmetric, in agreement with known characteristics of FWM due to the Kerr effect. Figure 3(b) shows the results of calculations with $\Omega$ on the F-SBS resonance $\Omega_7$ ($\Delta = 0$). In this case F-SBS leads to a significant transfer of power from the upper-frequency input pump wave $\vec{A}_1(z)$ to the lower-frequency one $\vec{A}_0(z)$, as also proposed in the analytic approximation. The coupling of power is associated with the stimulation of the guided acoustic mode. It also introduces pronounced asymmetry in the generation of sidebands: The first-order and second-order Stokes-wave components $\vec{A}_{-1,-2}(z)$ are amplified more efficiently than the corresponding anti-Stokes terms $\vec{A}_{2,3}(z)$. The calculations support the predicted general trends discussed earlier, and suggest that nonlinear wave mixing involving F-SBS is qualitatively different from a corresponding FWM process due to the Kerr effect alone.

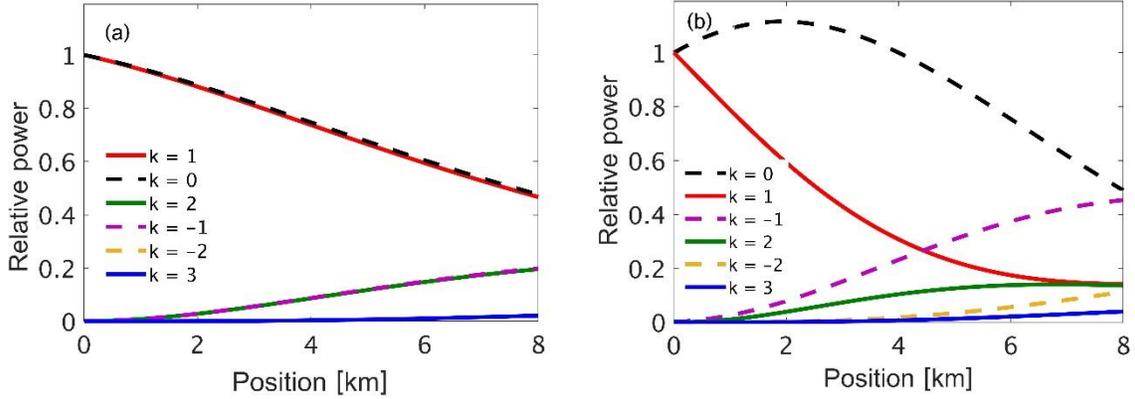

Fig. 3. Calculated local power levels $P_k(z)$ of six field components as functions of position along a standard single-mode fiber (see legends and Fig. 2(a) for notations). (a) The radio-frequency spacing $\Omega$ between the optical pump waves is detuned far from the F-SBS resonance, $\Delta = 8$. The power levels of the two input waves are indistinguishable, and so are the power levels of Stokes-wave and anti-Stokes wave sidebands of equal orders. (b) The spacing $\Omega$ is tuned to the F-SBS resonance $\Omega_7$ of guided acoustic mode $R_{0,7}$, $\Delta = 0$. A transfer of power takes place, from the upper-frequency pump wave to the lower frequency one. The nonlinear wave mixing amplification of sidebands becomes asymmetric.

Figure 4 compares between the approximate analytic solutions and the numerical results, for the two input waves and the first-order sidebands, with $\Delta = 0$. The simplified analytic expressions remain valid in the first 2-3 km, until the sidebands magnitudes become too large. As mentioned above, the analytic approximation cannot account for the asymmetry between Stokes and anti-Stokes sidebands.

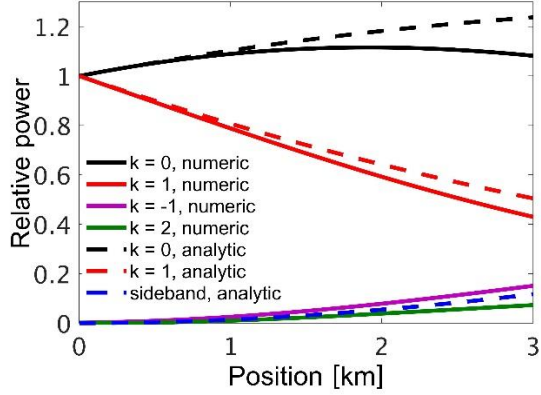

Fig. 4. Comparison between numerical simulations and approximate analytic solutions for local power levels of the two input pump waves and the two first-order sidebands (see legend). The radio-frequency spacing $\Omega$ between the optical pump waves is tuned to the resonance $\Omega_7$ of F-SBS through radial guided acoustic mode $R_{0,7}$, $\Delta = 0$.

Figure 5 shows the calculated power levels $P_k(z)$ for frequency offsets $\Omega$ that are detuned below or above the F-SBS resonance $\Omega_7$: $\Delta = \pm 1$ for panels (a) and (b), respectively. Although the magnitudes of the nonlinear opto-mechanical coefficients in Eq. (22) are the same in both cases, the nonlinear wave mixing dynamics are markedly different. The amplification of sidebands is larger with $\Omega$ detuned below the F-SBS resonance (Fig. 5(a)), in agreement with predictions (see Fig. 1). Figure 6 compares between numerical solutions and analytic approximations for the power levels of the two input wave components and the two first-order sidebands, with $\Delta = \pm 1$. Here too, the analytic expressions approximately account for the nonlinear wave mixing dynamics over the first 3 km. Note that the approximate solutions for $P_{1,0}(z)$, Eq. (11) and Eq. (12), do not distinguish between the signs of $\Delta$. The approximations are better for $\Delta = -1$ (Fig. 6(b)), since the amplification of sidebands is weaker in that case. The predictions of numerical simulations are tested against experimental results in the next section.

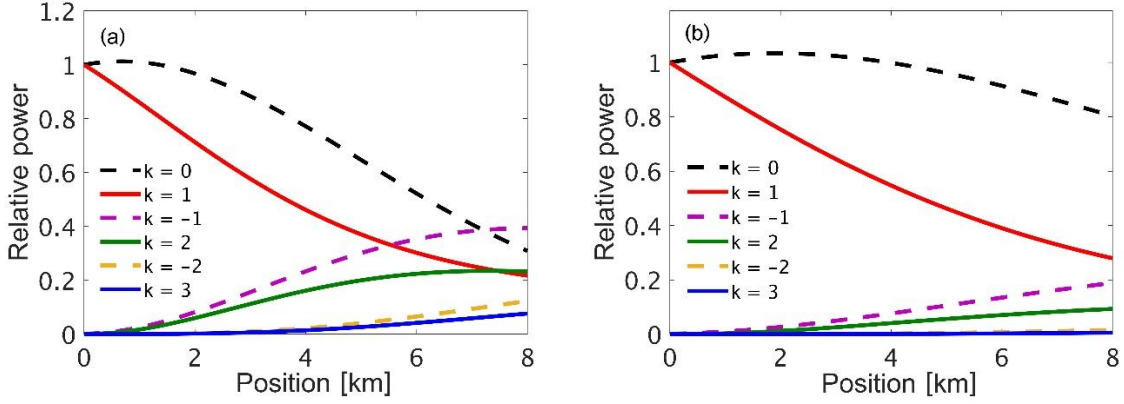

Fig. 5. Calculated local power levels $P_k(z)$ of six continuous optical waves as functions of position along a standard single-mode fiber (see legends and Fig. 2(a) for notations). Panel (a): The radio-frequency spacing $\Omega$ between the optical waves is detuned below the F-SBS resonance $\Omega_7$: $\Delta = 1$. Panel (b): The radio-frequency spacing $\Omega$ is detuned above $\Omega_7$: $\Delta = -1$. The nonlinear wave mixing amplification of sidebands is more pronounced with $\Omega$ detuned below resonance, panel (a), in agreement with expectations.

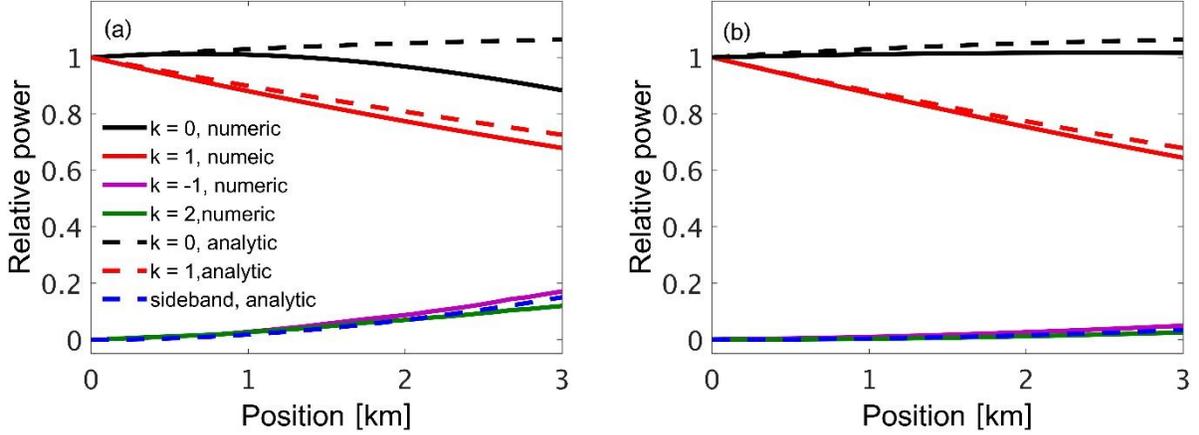

Fig. 6. Comparison between numerical simulations and approximate analytic solutions for local power levels of the two input pump waves and the two first-order sidebands (see legend of panel (a) for both panels). (a): The radio-frequency spacing $\Omega$ between the optical waves is detuned below the F-SBS resonance $\Omega_7$: $\Delta = 1$. Panel (b): The radio-frequency spacing $\Omega$ is detuned above $\Omega_7$: $\Delta = -1$.

## 3. Experimental setup, procedures and results

### *3.1 Experimental setup and procedures*

A schematic illustration of the multi-tone OTDR experimental setup used in the distributed mapping of FWM processes is shown in Fig. 7. Light from a tunable laser diode source of frequency $\omega$ at the 1550 nm wavelength range and 100 kHz linewidth was split in two paths. Light in one branch was used to generate the input wave components. It passed first through a double-sideband electro-optic modulator (EOM), which was biased for carrier suppression and driven by the output voltage of a sine-wave generator of variable radio-frequency $\frac{1}{2}\Omega$. The modulation generated two symmetric primary input tones at frequencies $\omega \pm \frac{1}{2}\Omega$. The two field components represent pump waves at frequencies $\omega_{1,0}$, respectively. The optical power of all higher-order modulation sidebands was at least 20 dB below those of the primary tones.

The two tones were amplified by an erbium-doped fiber amplifier (EDFA), and then amplitude-modulated by repeating, isolated square pulses in a second EOM. The duration $\tau$ of the pulses was 1 µs, and they were repeated every 125 µs. The pulses were launched into an 8 km-long fiber under test through a circulator. The peak power of the pulses at the input of the fiber under test was 160 mW (corresponding to $P_{1,0}(0) = 80$ mW each).

Nonlinear wave-mixing processes were mapped by reflectometry analysis. Rayleigh back-scatter from the fiber under test propagated back through the circulator into one end of a second fiber section of 180 m length, which served as a frequency-selective B-SBS amplifier. Light in the second output branch of the tunable laser diode source was used to generate the B-SBS pump wave. A suppressed-carrier, single-sideband EOM was employed to upshift the frequency of the laser diode light by an offset $\Delta\Omega_k = \Omega_B + (k - \frac{1}{2})\Omega$. Here $\Omega_B \sim 2\pi \times 10.72$ GHz is the Brillouin frequency shift of the B-SBS amplifier section. The B-SBS pump light was then amplified by a second EDFA to 150 mW power, and launched through a second circulator into the opposite end of the B-SBS amplifier section (see Fig. 7).

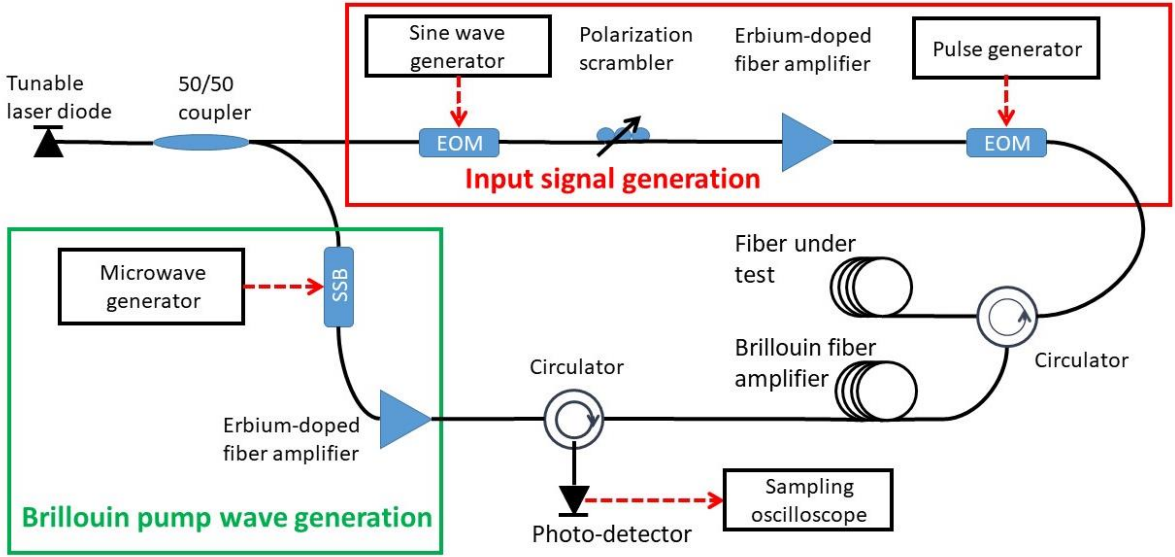

Fig. 7. Schematic illustration of the experimental setup used in the distributed analysis of nonlinear wave mixing along a fiber under test. EOM: electro-optic amplitude modulator; SSB: single-sideband electro-optic modulator.

Each choice of frequency offset $\Delta\Omega_k$ led to selective Brillouin amplification of the Rayleigh back-scatter of a specific tone, of optical frequency $\omega_k = \omega + \left(k - \tfrac{1}{2}\right)\Omega$, whereas back-scatter contributions at all other optical frequencies were unaffected. The B-SBS amplifier operated in the linear regime, with comparatively weak gain of 5 dB. Rayleigh backscatter contributions from different positions along the fiber under test are characterized by different states of polarization, and Brillouin gain is highly polarization-dependent [43]. The state of polarization of light at the input of the fiber under test was therefore scrambled, in order to obtain uniform Brillouin amplification of backscatter contributions from all locations.

The Rayleigh back-scatter trace at the output of the B-SBS amplifier was detected by a photo-receiver of 200 MHz bandwidth, and sampled by a real-time digitizing oscilloscope at 5 ns intervals. Traces were averaged over 128 repeating pulses and recorded for further offline processing. Following each acquisition, a reference trace was recorded with the B-SBS pump wave detuned to $\omega + 2\pi \times 12$ GHz. The reference trace consisted of unamplified back-scatter contributions, as well as residual leakage of the B-SBS pump. The reference was subtracted from each amplified data trace, thereby separating the amplified back-scatter contribution at the optical frequency of interest only.

The magnitude of Rayleigh back-scatter of coherent optical fields is extremely noisy [34,35]. For that reason, commercial OTDR instruments launch incoherent light pulses. However, the effective stimulation of guided acoustic waves requires the use of a narrowband, coherent light source. To work around this difficulty, experiments were repeated over 512 choices of the central optical frequency $\omega$, between wavelengths of 1558 nm and 1560 nm, and traces were averaged with respect to $\omega$ [36]. The ensemble average over many central optical frequencies helps reduce noise due to coherent Rayleigh back-scatter, and provides a mapping of local power levels along the fiber under test, similar to that of incoherent OTDR. As a final processing stage, the experimental traces were digitally filtered by a moving average window of duration $\tau$. The spatial resolution of the analysis is given by $\Delta z \approx \frac{1}{2} v_g \tau = 100$ m, where $v_g$ is the group velocity of light in the fiber. The experimental procedure was repeated for $k = -2, -1, 0, 1, 2, 3$, and for several choices of $\Omega$.

*3.2 Experimental results*

Figure 8(a) shows multi-tone OTDR measurements of the local power levels of the six spectral components $P_k(z)$, with $\Omega$ adjusted to $2\pi \times 340$ MHz. This radio-frequency is detuned by $2\pi \times 20$ MHz from the nearest resonance of F-SBS due to radial acoustic modes. Corresponding calculated traces are shown again for comparison (see Fig. 3(a)). Symmetric nonlinear generation of the Stokes-wave and anti-Stokes-wave sidebands is observed as anticipated. Figure 8(b) shows similar measurements and calculations with $\Omega$ adjusted to match $\Omega_7$: $2\pi \times 321.75$ MHz. Significant transfer or power from the upper-frequency pump wave to the lower-frequency one is observed, in excellent agreement with calculations. The parametric amplification of the Stokes-wave sidebands is stronger than that of the anti-Stokes-wave sidebands, as suggested by analysis and simulations. The experimental setup is sensitive and precise enough to resolve local power levels of individual tones below 1 mW, or about 1% of the input.

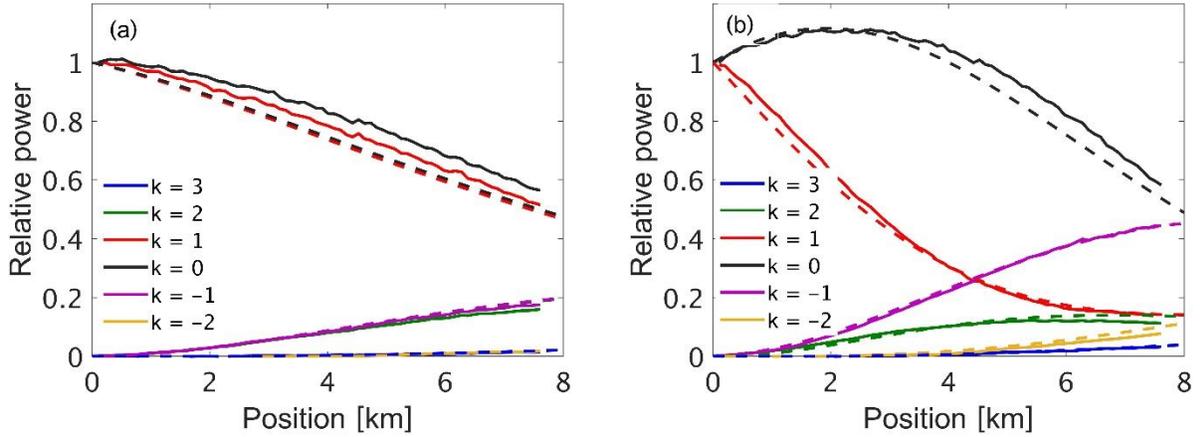

Fig. 8. Measured (solid traces) and calculated (dashed traces) local power levels $P_k(z)$ of six optical field components as functions of position along a standard single-mode fiber (see legends for colors). Panel (a): The radio-frequency spacing $\Omega$ between the input pump waves is detuned to $2\pi \times 340$ MHz, or $2\pi \times 20$ MHz from the nearest F-SBS resonances. Panel (b): The radio-frequency spacing $\Omega$ is adjusted to match the resonance frequency $\Omega_7$ of F-SBS through radial guided acoustic mode $R_{0,7}$: $2\pi \times 321.75$ MHz.

Figure 9(a) and Fig. 9(b) show the measured and calculated local power levels with $\Omega$ detuned below and above $\Omega_7$ by approximately half the F-SBS linewidth: $\Omega = 2\pi \times 319.25$ MHz and $\Omega = 2\pi \times 324.25$ MHz, respectively ($\Delta = \pm 1$). In both cases, asymmetry is observed between the two pump waves, as well as between the Stoke-wave and anti-Stokes-wave sidebands. The amplification of sidebands is more effective with $\Omega$ below the F-SBS resonance, as predicted. Here too, the agreement between measurements and simulations is very good.

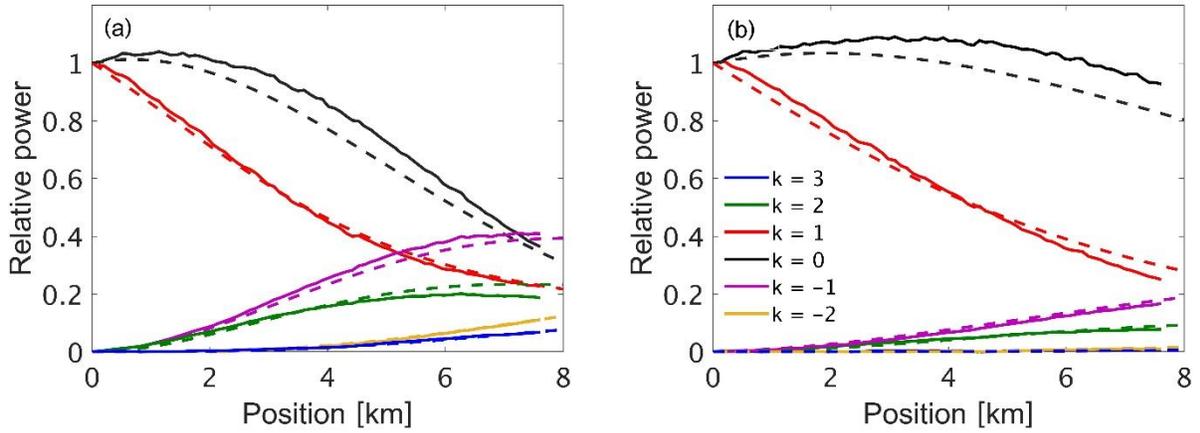

Fig. 9. Measured (solid traces) and calculated (dashed traces) local power levels $P_k(z)$ of six optical field components as functions of position along a standard single-mode fiber (see legend in panel (b) for colors). Panel (a): The radio-frequency spacing $\Omega$ between the input pump waves is detuned to $2\pi \times 319.25$ MHz, below the F-SBS resonance $\Omega_7$ ($\Delta = 1$). Panel (b): The radio-frequency spacing $\Omega$ is adjusted to $2\pi \times 324.25$ MHz, above the resonance frequency $\Omega_7$ ($\Delta = -1$).

## 4. Summary

Distributed mapping of nonlinear wave mixing processes involving F-SBS along standard single-mode fiber has been carried out in analysis, simulation and experiment. The results show that the combined effects of F-SBS and Kerr nonlinearity give rise to wave mixing processes that are markedly different from FWM that is driven by the Kerr effect alone. The process dynamics become strongly dependent on the exact frequency separation between co-propagating field components. When that separation is adjusted near a resonance of F-SBS, symmetry is removed. The stimulation of the guided acoustic wave is associated with coupling of power from a high-frequency input tone to a lower-frequency one. Consequently, Stokes-wave sidebands are amplified more efficiently than respective anti-Stokes-wave components. The detuning of the frequency separation below or above the F-SBS resonance leads to different wave mixing dynamics. The trends observed in the experiments are fully supported by corresponding numerical simulations. Approximate, analytic solutions for the regime of weak sidebands were derived and compared against simulations results.

The measurement protocol requires that the duration of pulses $\tau$ should be several times longer than the lifetime $\Gamma_m^{-1}$ of the guided acoustic mode. That lifetime is on the order of 100 ns in standard coated fibers [22]. This limitation restricts the spatial resolution $\Delta z$ of the analysis to the order of tens of meters or longer. Lifetime considerations also restrict the maximum input power of each pump wave, since the nonlinear length $L_{NL} = 1/\left[\gamma_{OM}^{(0,m)} P_{0,1}(0)\right]$ must not be shorter than $\Delta z$. This limitation, however, is not very stringent: input peak power levels of several W may still be used. The input power should also be kept below the threshold of amplified spontaneous backwards Brillouin scattering along the spatial extent of the pulse $2\Delta z$. This restriction is on the order of few W as well. Higher spatial resolution may be achieved using double-pulse acquisition schemes, as applied in distributed Brillouin sensing [44,45].

The mapping of F-SBS using Rayleigh back-scatter is at the basis of a new concept for distributed fiber-optic sensing, which we refer to as opto-mechanical time-domain reflectometry [36]. The protocol is based on Eq. (6) and Eq. (7). It relies on dual-tone OTDR measurements of power levels $P_{0,1}(z)$ to extract the local F-SBS spectra. The mechanical impedances of liquid media outside the fiber were successfully mapped based on estimates of the local F-SBS linewidth $\Gamma_m(z)$ [36]. Distributed sensing was demonstrated outside coated fibers as well [46]. The experimental procedure is similar to the one used here. However, as discussed in Section 2, the analysis of dual-tone traces is only valid when no other spectral field components exist in the fiber. The nonlinear wave mixing generation of spectral sidebands currently restricts the input power levels and the measurement signal-to-noise ratio, range and resolution [36].

On the other hand, the simultaneous OTDR analysis of four, six and perhaps more field components, as demonstrated in this work, might allow for the recovery of local F-SBS spectra even where nonlinear wave mixing takes place among multiple optical tones. The quantitative agreement between the coupled equations model (Eq. (22)) and experimental results suggests that such extensions of the sensing protocol may be feasible. The analysis of multiple sidebands might lead to higher resolution, longer range, and better precision. Sensing applications of multi-tone opto-mechanical time-domain reflectometry will be examined in future work.

**Acknowledgement**

The authors thank Prof. Mark Shtaif of Tel-Aviv University, Israel, for his advice on modeling of the Kerr effect in fiber. The authors acknowledge the financial support of the European Research Council (ERC), grant no. H2020-ERC-2015-STG 679228 (L-SID); and of the Israeli Ministry of Science and Technology, grant no. 61047.